\documentstyle[sprocl,epsfig]{article}

\def\NPB{{\em Nucl. Phys.} B}

\def\PRL{\em Phys. Rev. Lett.}
\def\PRD{{\em Phys. Rev.} D}

\def\be{\begin{equation}}
\def\ee{\end{equation}}
\def\bea{\begin{eqnarray}}
\def\eea{\end{eqnarray}}

\def\la{\hbox{{\lower -2.5pt\hbox{$<$}}\hskip -8pt\raise
-2.5pt\hbox{$\sim$}}}
\def\ga{\hbox{{\lower -2.5pt\hbox{$>$}}\hskip -8pt\raise
-2.5pt\hbox{$\sim$}}}

\begin{document}

\title{The Origin of the Highest Energy Cosmic Rays}

\author{A. V. OLINTO}

\address{Department of Astronomy \& Astrophysics, \\ \& Enrico Fermi
Institute, \\
The University of Chicago, Chicago, IL 60637, USA\\E-mail:
olinto@oddjob.uchicago.edu}

\maketitle\abstracts{ 
Contrary to expectations, several cosmic ray events with energies above
$10^{20}$ eV have been observed. The flux of such events is well above
the predicted Greisen-Zatsepin-Kuzmin cutoff due to the pion production
(via the $\Delta$ resonace) of extragalactic cosmic ray protons off the
cosmic microwave background. In addition to the relatively high flux of
events, the isotropic distribution of arrival directions and an
indication of hadronic primaries strongly challenge all models proposed
to resolve this puzzle. Models based on astrophysical accelerators need
to invoke strong Galactic and extragalactic magnetic fields with specific
properties which are yet to be observed, while models based on physics
beyond the standard model of particle physics generally predict photon
primaries contrary to experimental indications. The resolution of this
puzzle awaits a significant increase in the data at these energies which
will be provided by future experiments. }

\section{Introduction}

The role of cosmic rays in the discovery of fundamental properties of
nature dates back to the early 1900's. Over the last century,  Brazilian
physicists have actively contributed to these discoveries with
the most notable example of Prof. Cesar Lattes and the discovery of the
$\pi$ meson decay in Chacaltaya.\cite{lattes}  After a century of
direct and indirect cosmic ray detections, the data now spans 12 orders of
magnitude in energy  and about 32 in flux as shown in Figure
1.\cite{swordy} A strikingly simple power law spectrum over these many
decades in energy is observed: up to energies $\sim 10^{15}$ eV the
spectrum is $J(E)
\propto E^{-\gamma}$ with $\gamma \simeq 2.7$ and becomes steeper for
higher energies, $\gamma \simeq 3$. Most of these cosmic rays are 
believed to be accelerated in Galactic Supernovae shocks via Fermi
acceleration and to propagate diffusively in Galactic magnetic fields. 
This hypothesis is yet to be confirmed from direct observations of
Supernova remnants. In addition,  the energetic requirements at the
highest energies call for alternative sources which are expected to
originate from extragalactic sources since proton Larmor radii become
larger than Galactic dimensions at these extreme energies.  

\begin{figure}[thb]
 \begin{center}
  \mbox{\epsfig{file=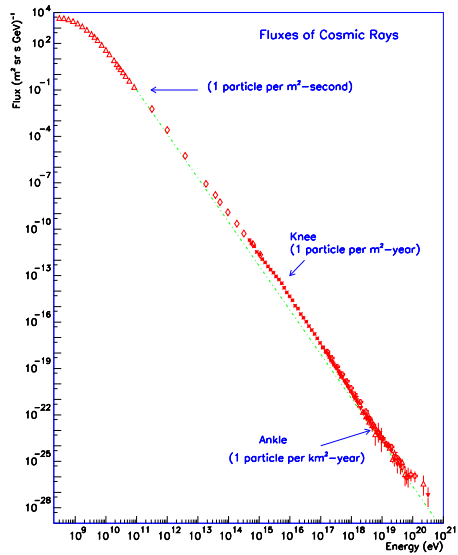,width=6cm}}
  \caption{\em {Cosmic ray spectrum compiled by S. Swordy [2].}}
 \end{center}
\end{figure}

Once extragalactic sources of protons are considered,  their
spectrum should not continue as a simple power law due to the effect of
the cosmic background radiation along their path. In the rest frame
of $10^{20}$ eV protons the cosmic microwave background 
corresponds to gamma rays with energies above the $\Delta$
resonance. Once protons reach these energies, they lose energy very
efficiently through the production of pions. As  
ultra-high energy protons travel across intergalactic space, the pion
photo-production process off the cosmic microwave background quickly
reduces their energy to below $\sim  7
\times 10^{19}$ eV as seen in Figure 2.\cite{cronin} If protons of
$10^{20}$ eV are observed, their sources are limited to distances of
less than  about $50\,$Mpc from Earth. This threshold energy loss process
should introduce a clear feature and a cutoff around
$\sim 10^{19-20}$ eV in the spectrum of extragalactic protons 
even if the injection spectrum at the source is a power law.  This
photopion production gives rise to the well-known 
Greisen-Zatsepin-Kuzmin\cite{GZK66} (GZK) cutoff  which is shown in Figure
3 by the dashed line for a uniform distribution of sources. 

\begin{figure}[thb]
 \begin{center}
  \mbox{\epsfig{file=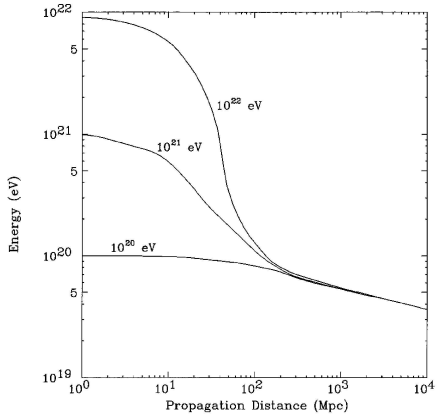,width=6cm}}
  \caption{\em {Proton energy vs. propagation distance
through the 2.7-K cosmic background radiation for the indicated
initial energies from  [2].
}}
 \end{center}
\end{figure}

Figure 3\cite{haya00} shows the expected spectrum  (flux multiplied by
$E^3$ versus energy) for extragalactic proton sources (dashed line) and
the flux  observed  by the Akeno Giant Air Shower Array (AGASA). AGASA
has accumulated many hundreds of events (728) with energies above
$10^{19}$ eV and 8 events above $10^{20}$ eV.\cite{haya00}
As can be seen from the figure, the data shows no indication of a cutoff.
This surprise has triggered considerable interest and the proposal of a
number of exotic scenarios designed to explain these data. In addition to
the ultra-high energy events detected by  AGASA, other experiments such as
Fly's Eye,  Haverah Park,  Yakutsk, and Volcano Ranch, have reported a
number of events at the highest energies  (see, e.g.,
[6] for reviews).

\begin{figure}[thb]
 \begin{center}
  \mbox{\epsfig{file=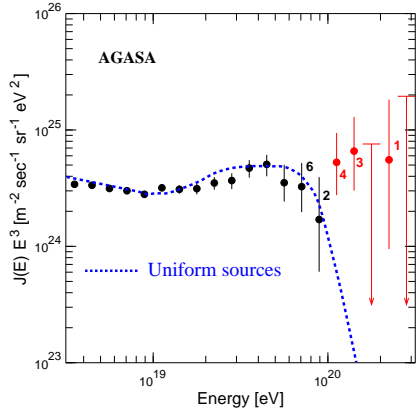,width=6cm}}
  \caption{\em {$E^3J(E)$ versus $E$ observed by AGASA and expected flux
for a uniform extragalactic source distribution [4]. }}
 \end{center}
\end{figure}

These observations are surprising because not only  the propagation of
particles at these energies is prone to large energy losses,  but the
energy requirements for astrophysical sources to accelerate particles to
$> 10^{20}$ eV are extraordinary. Different primaries do not
ease the difficulty. Ultra-high energy  nuclei  are
photodisintegrated on shorter distances due to the infrared background
\cite{PSB76SS99} while the radio background constrains photons to originate
from even closer systems.\cite{bere70PB96}

 The shape of the GZK
cutoff depends on the source input spectrum and the distribution of
sources in space as well as in the intergalactic magnetic field. In
Figure 4, we contrast the flux observed  by AGASA with a monte-carlo of
the expected flux for proton sources distributed homogeneously (shaded
region) or distributed like  galaxies (hatched region) with injection
spectrum $J(E) \propto E^{-\gamma}$ and $\gamma = 3$.\cite{BBO00} We
modeled the distribution of ultra-high energy cosmic ray (UHECR) sources
by using the galaxy distribution measured by the recent IRAS redshift
survey know as PSCz.\cite{saunders00a} As can be seen from the figure,
even allowing for the local overdensity the observations are
consistently above the theoretical expectation. In fact, when we
normalize our simulations by requiring that the number of events with $E
\ge 10^{19}$ eV equals the AGASA observations (728), we find that the
number of expected events for $E \ge 10^{20}$ eV is only $1.2 \pm 1.0$ for
the PSCz case, i.e., ``6 $\sigma$'' away from the observed 8 events.

 \begin{figure}[thb]
 \begin{center}
  \mbox{\epsfig{file=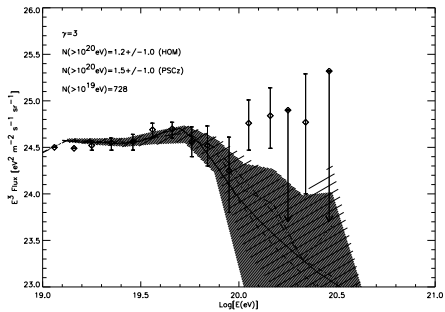,width=6cm}}
  \caption{\em {Simulated fluxes for the AGASA statistics of 728 events above 
$10^{19}$ eV, and $\gamma=3$, using a homogeneous source 
distribution ($\setminus$ hatches)
and the PSCz distribution (dense / hatches). The solid and dashed lines
are the results of the analytical calculations for the same two cases. 
The dash-dotted and dash-dot-dot-dotted lines trace 
the mean simulated fluxes for the homogeneous and the PSCz cases. 
(see [9]).}}
 \end{center}
\end{figure}

The gap between observed flux and model predictions narrows as the injection
spectrum of the UHECR sources becomes much 
harder than $\gamma = 3$.    For $\gamma = 2.1$ (shown in Figure
5\cite{BBO00}), the number of expected
events above $10^{20}$ eV reaches $3.3 \pm 1.6$ for a homogeneous
distribution while for the PSCz catalog it is $3.7\pm 2.0$.  This trend
can be seen also in Figure 6, where mean fluxes for $\gamma =1.5,
2.1$ and 2.7 are shown.

\begin{figure}[thb]
 \begin{center}
  \mbox{\epsfig{file=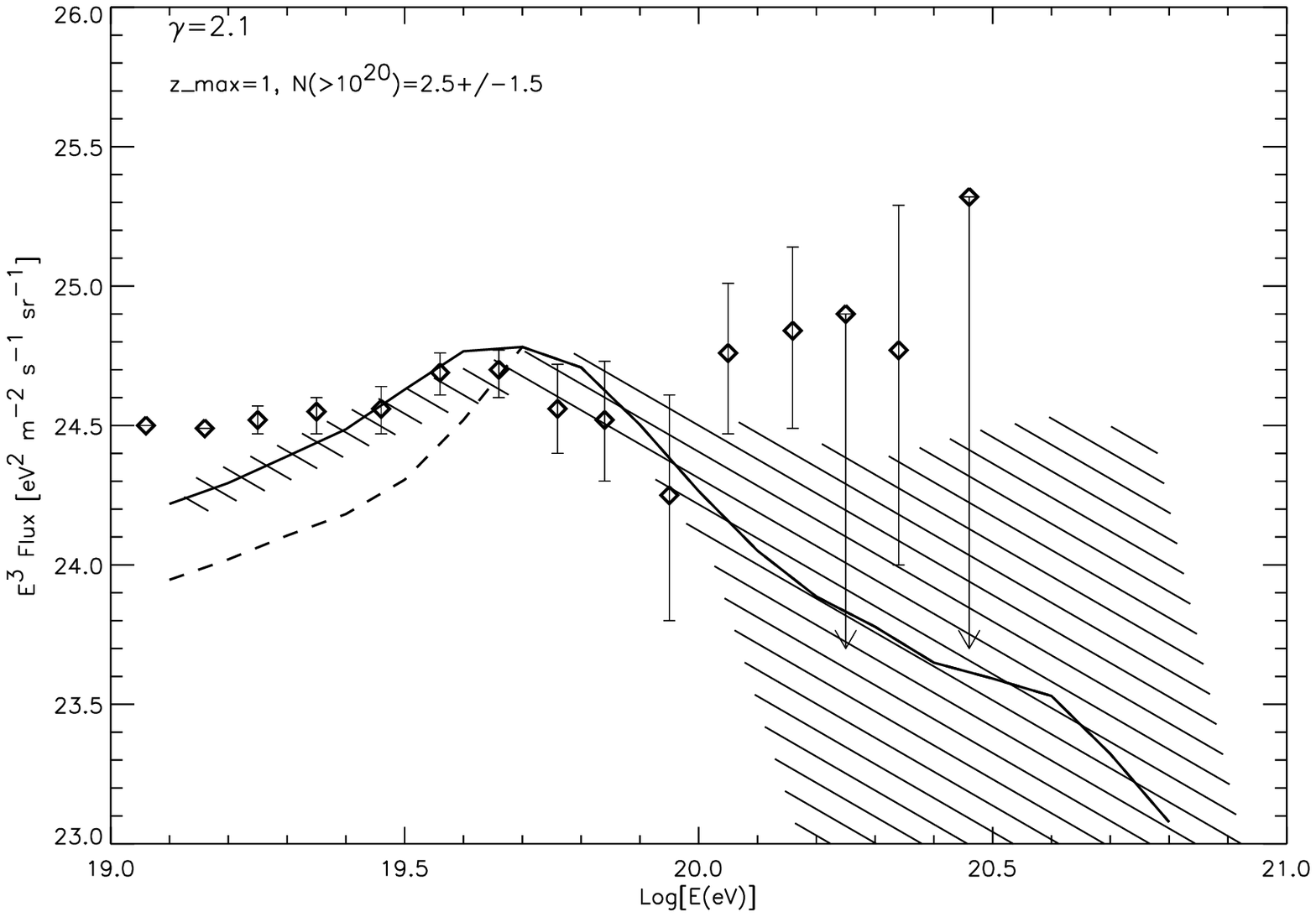,width=6cm}}
  \caption{\em {Simulated fluxes for the AGASA 
statistics of 728 events above 
$10^{19}$ eV, and $\gamma=2.1$, using a homogeneous source 
distribution with $z_{max}=0.1$ (/ hatches),
the PSCz distribution with $z_{max}=0.1$ (horizontal hatches), and
a homogeneous source 
distribution with $z_{max}=1$ ($\setminus$ hatches) (see [9]).
}}
 \end{center}
\end{figure}

In addition to the presence of events past the GZK cutoff, there has
been no clear counterparts identified in the  arrival direction of the
highest energy events. If these events are protons or photons,  these
observations should be astronomical, i.e., their arrival directions should
be the angular position of sources.  At these high energies the Galactic
and extragalactic magnetic fields should not affect proton orbits
significantly so that even protons would point back to their sources
within a few degrees. Protons at $10^{20}$ eV propagate mainly in
straight lines as they traverse the Galaxy since their gyroradii are
$\sim $ 100 kpc in $ \mu$G  fields which is typical in the Galactic disk.
Extragalactic fields are expected to be $\ll
\mu$G,\cite{KronVallee,BBO99} and induce at most  
$\sim$ 1$^o$ deviation from the source. Even if
the Local Supercluster has relatively strong fields, the highest energy
events may deviate at most $\sim$ 10$^o$.\cite{RKB98,SLB99}  At
present, no correlations between arrival directions and plausible optical
counterparts  such as sources in the Galactic plane, the Local Group, or
the Local Supercluster have been clearly identified.  Ultra high energy
cosmic ray data are consistent with an isotropic distribution of sources
in sharp contrast to the anisotropic distribution of light within 50 Mpc
from Earth.

The absence of a GZK cutoff and the isotropy of arrival directions are
two of the many challenges that models for the origin of UHECRs face.
This is an exciting open field, with many scenarios being proposed but 
no clear resolution.  Not only the origin of these 
particles  may be due to physics beyond the standard model of particle
physics, but their existence  can be used to constrain extensions of the
standard model such as violations of Lorentz invariance.\cite{ABGG00}

 \begin{figure}[thb]
 \begin{center}
  \mbox{\epsfig{file=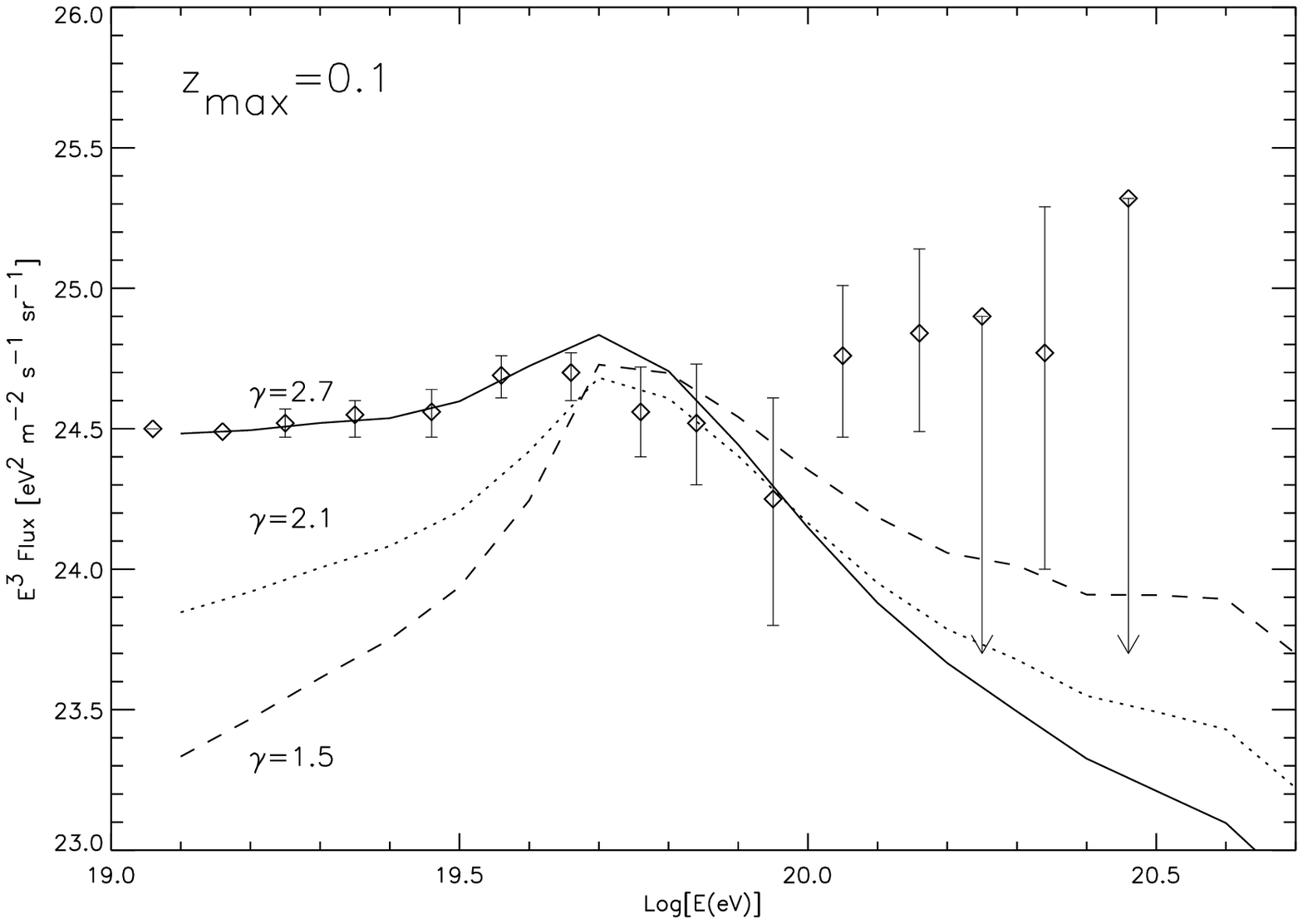,width=6cm}}
  \caption{\em {Propagated spectrum for source spectral index of 
$\gamma$ = 1.5, 2.1, 2.7}.}
 \end{center}
\end{figure}

In the next section, we discuss the issues involved in the propagation of UHECRs from
source to Earth. We then summarize the astrophysical Zevatron
proposals followed by the models that involve new physics. To
conclude, future observational tests of UHECR models and their
implications are discussed. Fore recent
reviews see [16-19].

\section{Propagation - Losses and Magnetic Fields}

Before contrasting plausible candidates for UHECR sources with the
observed spectrum and arrival direction distribution, we discuss the
effects of propagation from source to Earth.
Propagation studies involve both the study of losses along the primaries' path, such
as the photopion production responsible for the GZK cutoff,  as well as the
structure and magnitude of cosmic magnetic fields that determine the trajectories of
charged primaries and influence the development of the associated electromagnetic
cascade.\cite{LOS95PJ96}

For primary protons the main loss processes are photopion production off the CMB
that gives rise to the GZK cutoff\cite{GZK66} and  pair
production.\cite{Blu70}   For straight line propagation, the cutoff
should be present at $\sim 5 \times 10^{19}$ eV  and a significant number
of hard sources should be located within $\sim$ 50 Mpc from us. As we
discussed above, even with the small number of accumulated events at the
highest energies, the AGASA spectrum is incompatible  with a  GZK cutoff
for a homogeneous extragalactic source distribution.  The shape of the
cutoff can be modified if the distribution of sources is not
homogeneous\cite{BBO00,BBDGP90,BW00} and if  the particle trajectories
are not rectilinear (e.g., the case of sizeable intergalactic magnetic
fields).\cite{WME96}$^-$\cite{BO99,SLB99}

Charged particles of energies up to $10^{20}\,$eV can be deflected
significantly in cosmic magnetic fields. In a constant magnetic field
of strength $B= B_6 \mu$G, particles of energy $E= E_{20}10^{20}{\rm
eV}$ and charge $Ze$ have Larmor radii of
$r_L \simeq 110$ kpc $(E_{20}/B_6 Z)$.  If the UHECR primaries are
protons, only large scale intergalactic magnetic fields affect their
propagation significantly unless
the Galactic halo has  extended fields.\cite{S97} For higher $Z$, the
Galactic magnetic field can strongly affect the trajectories of
primaries.\cite{ZPPR98,HMR99}

 Whereas Galactic magnetic fields are reasonably
well studied, extragalactic fields are still very
ill understood.\cite{KronVallee} Faraday rotation measures show large
magnitude fields   ($\ga \,
\mu$G) in the central regions  of clusters of
galaxies. In regions between clusters, the presence of
magnetic fields is evidenced by synchrotron emission but the strength
and structure are yet to be determined. On the largest
scales, limits can be imposed by the observed isotropy of the CMB
and by a statistical interpretation of Faraday rotation measures of light
from distant quasars. The isotropy of the CMB can 
constrain the present horizon scale fields
$B_{H_0^{-1}} \la \, 3 \times 10^{-9}$ G.\cite{BFS97} Although the
distribution of Faraday rotation measures have large non-gaussian tails,
a reasonable limit can be derived using the  median of the  distribution
in an inhomogeneous universe: for fields assumed to be constant on the
present horizon scale, $B_{H_0^{-1}} \la \, 10^{-9}$ G; for fields with 50
Mpc coherence length,  $B_{50 {\rm Mpc}}
\la \, 6 \times 10^{-9}$ G; while for  1 Mpc coherence length, $B_{Mpc}
\la \, 10^{-8}$ G.\cite{BBO99} These limits apply to a $\Omega_b h^2 =
0.02$ universe and use quasars up to redshift $z=2.5$.  Local
structures can have fields above these upper limits as long as they are
not common along random lines of site between
$z$ = 0 and 2.5.\cite{RKB98,BBO99}

Of particular interest is the field in the local 10 Mpc volume around
us.  If the Local Supercluster has fields of about
$10^{-8}$ G or larger, the propagation of ultra high energy protons
could become diffusive and the spectrum and angular distribution at the
highest energies would be significantly
modified.\cite{RKB98,BO99,WW79GWW80BGD89}  For example,  in Figure
7,\cite{BO99} a source with spectral index 
$\gamma\ga 2$ that can reach $E_{max} \ga 10^{20}$ eV is constrained by
the overproduction of lower energy events around 1 to 10 EeV (EeV $\equiv
10^{18}$ eV). Furthermore, the structure
and magnitude of magnetic fields in the Galactic halo\cite{S97,HMR99} or
in a possible Galactic wind  can also affect the observed UHECRs. In
particular, if our Galaxy has a strong magnetized wind, what appears to
be an isotropic distribution  in arrival directions may have
originated on a small region of the sky such as the Virgo
cluster.\cite{ABMS99} In the future, as sources of UHECRs are
identified,  large scale magnetic fields will be better
constrained.\cite{LSOS97}

\begin{figure}
 \mbox{\epsfig{file=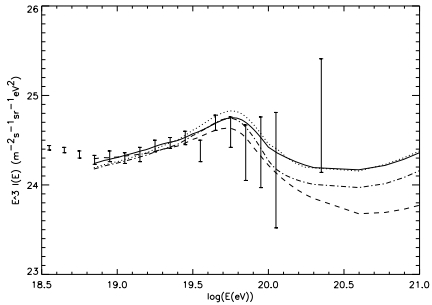,width=6cm}}
\caption{Flux vs. Energy with $E_{max} = 10^{21}$ eV at source. Choices
of source distance r(Mpc), spectral index $\gamma$,  proton
luminosity $L_p$(erg/s), and LSC field B($\mu$G) are: solid line (13
Mpc,  2.1, $2.2 \times 10^{43}$ erg/s, 0.05 $\mu$G); dotted line (10
Mpc,  2.1, $10^{43}$ erg/s, 0.1 $\mu$G); dashed line (10 Mpc, 2.4, 3.2
$\times 10^{43}$ erg/s, 0.1 $\mu$G);  and dashed-dotted line (17 Mpc, 
2.1, $3.3 \times 10^{43}$ erg/s, 0.05 $\mu$G).} 
\end{figure}

If cosmic rays are heavier nuclei, the attenuation length is shorter 
than that for protons due to photodisintegration on the infrared
background.\cite{PSB76SS99} However, UHE nuclei may be of Galactic
origin. For large enough charge, the trajectories of UHE nuclei
are significantly affected by the Galactic magnetic field\cite{HMR99}
such that a Galactic origin can appear isotropic.\cite{ZPPR98} The
magnetically induced distortion of the flux map   of UHE events can give
rise to some higher flux regions where caustics form and some much lower
flux regions (blind spots).\cite{HMR99} Such propagation effects are one
of the reasons why full-sky coverage is necessary for resolving the UHECR
puzzle.

The trajectories of photon primaries are not affected by magnetic
fields, but energy losses due to the radio background
constrain photons to originate from systems at
$\la 10\,$Mpc.\cite{S69CBA70bere70PB96} If associated with
luminous systems, sources of UHE photons should point back to their
nearby sources. The lack of counterpart identifications suggests that if
the primaries are photons, their origin involves
physics beyond the standard model.

\section{Astrophysical Zevatrons}

The puzzle presented by the observations of cosmic rays above $10^{20}$ eV
have generated a number of  proposals that we divide here as {\it
Astrophysical Zevatrons} and {\it New Physics} models.  Astrophysical
Zevatrons are also referred to as  bottom-up models and involve searching
for acceleration sites in known astrophysical objects that can reach ZeV
energies. New Physics proposals can be either hybrid or pure top-down
models. Hybrid models involve Zevatrons and extensions of the particle
physics standard model while top-down models involve the decay of very
high mass relics from the early universe and physics way beyond the standard
model. Here we discuss astrophysical Zevatrons while new physics models
are discussed in the next section. 

Cosmic rays can be accelerated in
astrophysical plasmas when large-scale macroscopic motions, such
as shocks, winds, and turbulent flows, are transferred to individual
particles. The maximum energy of accelerated  particles,
$E_{\rm max}$, can be estimated by requiring that the gyroradius of the
particle be contained in the acceleration region: $E_{\rm max} = Ze \, B
\, L$, where  $Ze$ is the charge of the particle, $B$ is the strength  and
$L$ the  coherence length of the magnetic field embedded in the plasma.
For $E_{\rm max} \ga 10^{20}$ eV and $Z \sim 1$, the only known
astrophysical sources with reasonable  $B L $ products   are neutron
stars ($B \sim 10^{13}$ G, $L \sim 10$ km),  active galactic nuclei (AGNs)
($B \sim 10^{4}$ G, $L \sim 10$ AU), radio lobes of AGNs ($B \sim
0.1\mu$G, $L \sim 10$ kpc), and clusters of galaxies ($B \sim \mu$G, $L
\sim 100$ kpc). In Figure 8, we highlight the  $B$ vs. $L$  for objects
that can reach  $E_{max} =  10^{20}$ eV with $Z=1$ (dashed line) and
$Z=26$  (solid line).   We discuss each of these candidates below.

\begin{figure}[htb]
  \mbox{\epsfig{file=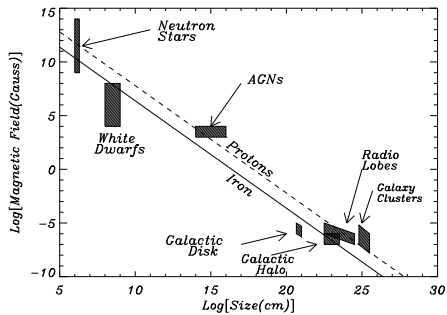,width=6cm}}
  \caption{\em {$B$ vs. $L$, for $E_{max} =  10^{20}$ eV, $Z=1$
(dashed line) and $Z=26$  (solid line) from [16].}}
\end{figure}

{\it Clusters of Galaxies:}
Cluster shocks are
reasonable sites to consider for UHECR
acceleration, since  particles with energy up to $E_{\rm max}$ can be
contained by cluster fields. However, efficient losses due to 
photopion production off the CMB during the propagation inside the
cluster limit UHECRs in cluster shocks  to reach at most
$\sim$ 10 EeV.\cite{KRJ96KRB97}

{\it AGN Radio Lobes:}
Next on the list of plausible Zevatrons
are extremely powerful radio galaxies.\cite{BS87B97}
Jets from the central black-hole of an active galaxy end at a termination
shock where the interaction of the jet with the intergalactic medium
forms radio lobes and  `hot spots'. Of special interest are the most
powerful AGNs where shocks can accelerate particles to energies well
above an EeV via the first-order Fermi mechanism. These sources may be
responsible for the flux of UHECRs up to the GZK cutoff.\cite{RB93}

A nearby specially powerful source may be able to reach energies  past the
cutoff.  However, extremely powerful AGNs with radio lobes
and hot spots are rare and far apart. The closest known object is M87 in
the Virgo cluster ($\sim$ 18 Mpc away)  and could be a main source of
UHECRs.  Although a single nearby source with especially hard spectra may 
fit the  spectrum  for a given strength and structure of the intergalactic
magnetic field,\cite{BO99} it is unlikely to match the observed arrival
direction distribution. If M87 is the primary source of UHECRs a
concentration of events in the direction of M87 or the Virgo cluster
should be seen in arrival directions. No such hot spot is
observed ({\it Hot spot} in Table 1). The next known nearby
source after M87 is NGC315 which is already too far at a distance of
$\sim $ 80 Mpc. Any unknown source between M87 and NGC315  would likely
contribute a second hot spot, not an isotropic distribution. The very
distant radio lobes will contribute a GZK cut spectrum which is
not observed.

The lack of a clear hot spot in the direction of M87 has encouraged the
idea that a strong Galactic magnetic wind may exist that could help
isotropize the arrival directions of UHECRs.  A  Galactic wind with a 
strongly magnetized azimuthal component\cite{ABMS99} ($Galactic Wind$ in
Table 1) can significantly alter the paths of UHECRs such that the
observed arrival directions of events above 10$^{20}$ eV would trace
back to the North Galactic Pole which is close to the Virgo where M87
resides. If our Galaxy has such a wind is yet to be determined. The
proposed wind would focus most observed events into the
northern Galactic pole and render point source identification
fruitless.\cite{BLS00} Future observations of UHECRs from the Southern
Hemisphere   by the Southern Auger Observatory will provide precious data
on previously unobserved parts of the sky  and help   distinguish
plausible proposals for the effect of local magnetic fields on arrival
directions.  Full sky coverage is a key discriminator of such
proposals.   

{\it  AGN - Central Regions:}
The powerful engines that give rise to the observed jets and radio
lobes are located in the central regions of active galaxies and are
powered by the accretion of matter onto supermassive black holes. It
is reasonable to consider the central engines themselves as the likely
accelerators.\cite{T86} In principle, the nuclei of  generic
active galaxies (not only the ones with radio lobes) can accelerate
particles via a unipolar inductor not unlike the one operating in
pulsars. In the case of AGNs,   the magnetic field  may be provided by
the infalling matter and the spinning black hole horizon provides the
imperfect conductor for the unipolar induction. 

The problem with AGNs as UHECR sources is two-fold: first, UHE particles
face  debilitating losses in the acceleration region due to the intense
radiation field present in AGNs,  and second, the
spatial distribution of objects should give rise to a  GZK cutoff of the
observed spectrum. In the central  regions of AGNs, loss processes are
expected to downgrade particle energies well below the maximum
achievable energy. This limitation has led to the proposal that  quasar
remnants, supermassive black holes in centers of inactive galaxies,  are
more effective UHECR accelerators.\cite{BG99} In this case, losses at the
source are not as significant but the propagation from source to us should
still lead to a clear GZK cutoff since sources would be associated with the
large scale structure of the galaxy distribution ({\it LSS} in Table 1).
From Figure 4--6, these models can only succeed if  the source spectrum
is fairly hard ($\gamma \la 2$).\cite{BBO00}

\begin{table*}[hbt]
\setlength{\tabcolsep}{1.1pc}
\newlength{\digitwidth} \settowidth{\digitwidth}{\rm 0}
\catcode`?=\active \def?{\kern\digitwidth}
\caption{Zevatrons}
\label{tab:zevatrons}
\medskip
\begin{tabular*}{\textwidth}{@{}l@{\extracolsep{\fill}}rrrr}
                 & \multicolumn{1}{r}{Radio Lobes} 
                 & \multicolumn{1}{r}{AGN Center} 
                 & \multicolumn{1}{r}{YNSWs}         
                 & \multicolumn{1}{r}{GRBs}         \\
\hline
Composition & Proton &  Proton & Iron & Proton  \\  
Source $\gamma$ & 2--3 & 2--3 & 1  & $\Delta E/E \sim 1$ 
 \\ Sky Distrib. & M87(1 spot) & LSS & Galaxy & Hot Spot
\\ $B$ Needs & Galactic Wind & -- & $B_{gal}$  &  $B_{IGM}$  
\\ Difficulties  & Hot Spot & GZK  & Iron  & flux,
$B_{IGM}$ \\
\hline
\end{tabular*}
\end{table*}

 {\it  Neutron Stars:}
Another astrophysical system capable of accelerating UHECRs is a neutron
star. In addition to having the ability to confine $10^{20} eV$ protons
(Figure 8), the rotation energy of young neutron stars is more than
sufficient to match the observed UHECR fluxes.\cite{VMO97}
However, acceleration processes inside the neutron star light cylinder are
bound to fail much like the AGN central region case:  ambient magnetic
and radiation fields induce significant losses. However, the plasma that
expands beyond the light cylinder is free from the main loss processes
and may be accelerated to ultra high energies.

One possible source of UHECR past the GZK cutoff is the early evolution
of neutron stars. In particular, newly formed, rapidly rotating
neutron stars may accelerate iron nuclei  to UHEs  through relativistic
MHD winds beyond  their light cylinders.\cite{BEO99} This mechanism
naturally leads to very hard injection spectra ($\gamma \simeq 1$) (see 
Table 1). As seen in Figure 6,   $\gamma \sim 1$ improves the agreement
between predicted flux and observations for energies above $10^{20}$ eV. In
this case, UHECRs originate mostly in the Galaxy and the arrival directions
require that the primaries be  heavier nuclei. Depending on the structure
of   Galactic magnetic fields, the trajectories of iron nuclei from Galactic
neutron stars may be consistent with the observed arrival directions of the
highest energy events.\cite{ZPPR98} Moreover,  if  cosmic rays  of a few
times $10^{18}$ eV are protons of Galactic origin, the isotropic
distribution observed at these energies is indicative of the diffusive
effect of the Galactic magnetic fields on iron at $\sim 10^{20}$ eV. This
proposal should be constrained once the primary composition is clearly
determined (see {\it Iron} in Table 1).

It has also been suggested that young extragalactic highly
magnetized neutron stars (magnetars) may be sources of UHE protons which
are accelerated by reconnection events.\cite{GL00} These would be prone
to a GZK cut spectrum and would need a very hard injection spectrum to
become viable explanations. 

{\it  Gamma-Ray Bursts:}
Transient high energy phenomena such as gamma-ray
bursts (GRBs) may also be a source of ultra-high energies
protons.\cite{WV95} In addition to both phenomena having unknown origins,
GRBs and UHECRs have other similarities that may argue for a common
source. Like UHECRs, GRBs are distributed isotropically in the sky,  and
the average rate of $\gamma$-ray energy emitted by GRBs is comparable to
the energy generation rate of UHECRs of energy $>10^{19}$ eV in a
redshift independent cosmological distribution of sources, both have  $ 
\approx 10^{44}{\rm erg\ /Mpc}^{3}/{\rm yr}.$ 

However, recent GRB counterpart
identifications argue for a strong cosmological evolution for GRBs. 
 The redshift dependence of GRB distribution is such that the flux of
UHECR associated with nearby GRBs would be too small to fit the
UHECR observations.\cite{Ste99} In addition, the distribution of UHECR
arrival directions and arrival times argues against the GRB--UHECR common
origin. Events past the GZK cutoff require that only GRBs from $\la 50$
Mpc contribute. Since less than about {\it one} burst is expected to have
occurred within this region over a period of 100 yr, the unique source
would appear as a concentration of UHECR events in a small part of
the sky (a {\it Hot spot} in Table 1).   In addition, the signal would be
very narrow in energy  $\Delta E/E\sim1$. Again, a strong intergalactic
magnetic field can ease some of these difficulties giving a very
large dispersion in the arrival time and direction of protons  produced in
a single burst  ({\it large} $B_{IGM}$ in Table 1).\cite{WV95}   Finally,
if the observed small scale clustering of arrival directions is confirmed
by future experiments with clusters having some lower energy events
clearly precede higher energy ones, bursts would be
invalidated.\cite{SLO97}

\section{New Physics Models}

The UHECR puzzle has inspired a number of different models that involve
physics beyond the standard model of particle physics. New Physics
proposals can be top-down models or a hybrid of astrophysical Zevatrons
with new particles. Top-down models involve the decay of very high mass
relics that could have been formed in the early universe.

The most economical among hybrid proposals involves a familiar
extension of the standard model, namely, neutrino masses.  If
some flavor of neutrinos have mass (e.g., $\sim 0.1 eV$), the relic
neutrino background  is a target for extremely high energy neutrinos to
interact and generate other particles by forming a Z-boson that
subsequently decays\cite{We97FMS97} (see $\nu$ {\it Z burst} in Table
2). If the universe has very luminous sources (Zevatrons) of extremely
high energy neutrinos ($\gg 10^{21}$ eV), these neutrinos would traverse
very large distances before annihilating with neutrinos in the smooth
cosmic neutrino background.  UHE
neutrino Zevatrons can be much further than the GZK limited volume,
since neutrinos do not suffer the GZK losses. But if the interaction
occurs throughout a large volume, the GZK feature should also be
observed.  For plausible neutrino masses $\sim 0.1 eV$, the neutrino
background is very unclustered, so the arrival direction for events
should be isotropic and small scale clustering may be a strong challenge
for this proposal.  The weakest link in this proposal is the nature of a
Zevatron powerful enough to accelerate protons above tens of ZeVs that
can produce  neutrinos as secondaries with $\ga$ ZeV. This Zevatron is quite
spectacular, requiring an energy generation in excess of presently known
highest energy sources (referred to as {\it
ZeV $\nu$'s} in Table 2).

Another suggestion is that the UHECR primary is a new hadronic
particle that is also accelerated in Zevatrons. The mass of a hypothetical
hadronic primary can be limited by the shower development of the Fly's
Eye highest energy event to be below $\la 50$ GeV.\cite{AFK98}  As in
the Z-burst proposal, a neutral particle  is usually harder to accelerate
and are usually created as secondaries of even higher energy charged
primaries. But once formed these can traverse large distances without
being affected by cosmic magnetic fields. Thus, a signature for future
experiments  of hybrid models that invoke new particles as primaries is a
clear correlation between the position of powerful Zevatrons in the sky
such as distant compact radio quasars and the arrival direction of  UHE
events.\cite{FB98} Preliminary evidence for such a correlation has been
recently reported.\cite{VBJRRM}       

Another exotic primary that can be accelerated to ultra high energies by
astrophysical systems is the vorton. Vortons are small loops of
superconducting cosmic string stabilized by the angular momentum of
charge carriers.\cite{DS89} Vortons can be a component of the dark
matter in galactic halos and be accelerated by astrophysical magnetic
fields.\cite{BP97}  Vortons as primaries can be constrained by the
observed shower development profile.

It is possible that none of the astrophysical scenarios or the hybrid new
physics models are able to explain present and  future UHECR
data. In that case,  the alternative is to consider top-down models.
Top-down models involve  the decay of monopole-antimonoploe
pairs,\cite{H83HS83}  ordinary and superconducting cosmic strings,  cosmic
necklaces,  and superheavy long-lived relic particles.  The idea
behind these models is that relics of the very early universe,
topological defects (TDs) or superheavy relic (SHR) particles,  produced 
after or at the end of inflation, can decay today and generate UHECRs. 
Defects, such as cosmic strings, domain walls, and magnetic monopoles, 
can be generated through the Kibble mechanism  as symmetries are broken
with the expansion and cooling of the universe.  Topologically stable
defects can survive to the  present and decompose into their constituent
fields  as they collapse,  annihilate, or reach critical current in the
case of superconducting cosmic strings. The decay products, superheavy
gauge and higgs bosons, decay into jets of hadrons, mostly pions.  Pions
in the jets subsequently decay into $\gamma$-rays, electrons, and
neutrinos. Only a few percent of the hadrons are expected to be nucleons.
Typical features of these scenarios are a predominant release of
$\gamma$-rays and neutrinos and a QCD fragmentation spectrum which is 
considerably harder than the case of Zevatron shock acceleration.  

ZeV energies are not a challenge for top-down models since symmetry
breaking scales at the end of inflation typically are $\gg 10^{21}$
eV (typical X-particle masses vary between 
$\sim 10^{22-25}$ eV).  Fitting the observed flux
of UHECRs is the real challenge since the typical distances between TDs
is  the  Horizon scale, i. e., about several Gpc. The low flux hurts proposals
based on ordinary  and superconducting cosmic strings which are
distributed throughout  space ({\it Extragal. TD} in Table 2).
Monopoles usually suffer the opposite problem, they would in general be
too numerous. Inflation succeeds in diluting the number density of
monopoles  and makes them too rare for UHECR production. To reach
the observed UHECR flux, monopole models usually involve some degree of
fine tuning. If enough monopoles and antimonopoles survive from the early
universe,  they may form a bound state, named monopolonium, that  can 
decay  generating UHECRs. The lifetime of monopolonia may be too short
for this scenario to succeed unless they are connected by
strings.\cite{PO99}

\begin{figure}[thb]
 \begin{center}
  \mbox{\epsfig{file=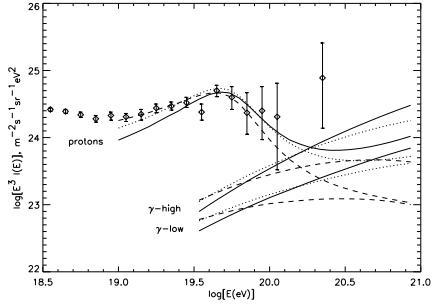,width=6cm}}
  \caption{\em {Proton and $\gamma$-ray fluxes from necklaces for 
$m_X=  10^{14}$ GeV (dashed lines),  $10^{15}$ GeV (dotted 
lines), and $10^{16}$ GeV (solid lines) normalized to
the  observed data.
$\gamma$-high  and  $\gamma$-low  correspond to two extreme cases 
of $\gamma$-ray absorption (see, [56]).
}}
\end{center}
\end{figure}

Once two symmetry breaking scales are invoked, a combination of
horizon scales gives room to reasonable number densities. This can be
arranged for cosmic strings that end in monopoles making a monopole
string network or even more clearly for cosmic necklaces.\cite{BV97}
Cosmic necklaces are hybrid defects where each monopole is connected to
two strings resembling beads on a cosmic string necklace. Necklace
networks may evolve to configurations that can fit the UHECR flux which
is ultimately generated by the annihilation of monopoles with
antimonopoles trapped in the string.\cite{BV97,BBV98}  In these
scenarios, protons dominate the flux in the lower energy side of the GZK
cutoff  while photons tend to dominate at higher energies depending on
the radio background (see Figure 9 and {\it Extragal. TD} in Table 2).
If  future data can settle the composition of UHECRs from 0.01 to 1 ZeV,
these models can be well constrained. In addition to fitting the UHECR
flux, topological defect models are constrained by limits on the flux
of  high energy photons, from 10 MeV to 100 GeV, observed by EGRET.

\begin{figure}[thb]
 \begin{center}
  \mbox{\epsfig{file=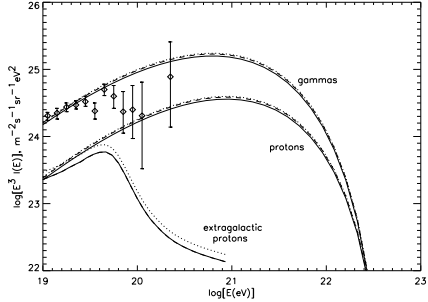,width=6cm}}
  \caption{\em {SHRs  or monopolia decay fluxes
(for $m_X= 10^{14} ~GeV$):
nucleons from the halo ({\it protons}), $\gamma$-rays
from the halo ({\it gammas}) and extragalactic protons. Solid, dotted
and dashed curves correspond to different  model parameters
(see [56]).
}}
\end{center}
\end{figure}

Another interesting possibility is the recent proposal that UHECRs are
produced by the decay of unstable superheavy relics that live much longer
than the age of the universe.\cite{BKV97}   SHRs may be produced at
the end of inflation by non-thermal effects such as a varying
gravitational field, parametric resonances during preheating,  instant
preheating, or the decay of topological defects. 
These models need to invoke special symmetries to insure unusually long
lifetimes for SHRs and that a sufficiently small percentage decays today
producing UHECRs.\cite{BKV97,CKR99KT99}  As in the topological defects
case, the decay of these relics also generates jets of hadrons.  These
particles behave like cold dark matter and could constitute a fair
fraction of the halo of our Galaxy. Therefore, their halo decay products
would not be limited by the GZK cutoff allowing for a large flux at UHEs
(see Figure 10 and {\it SHRs} in Table 2). Similar signatures can occur
if topological defects are microscopic, such  as monopolonia and
vortons, and decay in the Halo of our Galaxy ({\it Halo TD} in Table
2). In both cases ({\it SHRs} and {\it Halo TD}) the composition of the
primary would be a good discriminant since the decay products are
usually dominated by photons.

\begin{table*}[hbt]
\setlength{\tabcolsep}{1.1pc}
\caption{NEW PHYSICS}
\label{tab:newphys}
\medskip
\begin{tabular*}{\textwidth}{@{}l@{\extracolsep{\fill}}rrrr}
                 & \multicolumn{1}{r}{$\nu$ Z burst} 
                 & \multicolumn{1}{r}{Extragal TD} 
                 & \multicolumn{1}{r}{Halo TD}         
                 & \multicolumn{1}{r}{SHRs}         \\
\hline
Composition & Photon &  Pho+GZK p & Photon & Photon  \\  
Source $\gamma$ & Z frag & QCD frag & QCD frag  & QCD frag
 \\ Sky Distrib. & Isotropic & Isotropic & Gal Halo & Gal Halo
\\ Problem:Pho+  & ZeV $\nu$'s &  flux  & origin  &
lifetime
\\
\hline
\end{tabular*}
\end{table*}

Future experiments should be
able to probe these hypotheses.  For instance, in the case of SHR
and monopolonium decays, the arrival
direction distribution should be close to isotropic but show an
asymmetry due to the position of the Earth in the Galactic
Halo\cite{BBV98} and the clustering due to small scale dark matter
inhomogeneities.\cite{BlSe00} Studying plausible halo models for
their expected asymmetry and inhomogeneitis will help constrain halo
distributions especially when larger data sets are available in the future.
High energy gamma ray experiments such as GLAST will also help constrain 
SHR models via the electromagnetic decay products.\cite{B99}

\begin{figure}[thb]
 \begin{center}
  \mbox{\epsfig{file=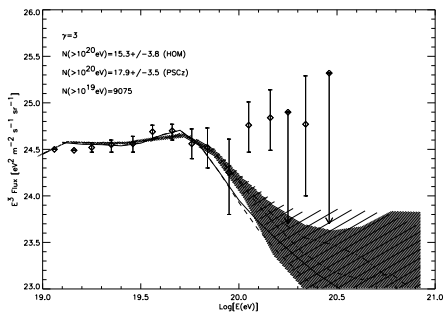,width=6cm}}
  \caption{\em {Simulated fluxes for the Auger projected 
statistics of 9075 events above 
$10^{19}$ eV, and $\gamma=3$, using a homogeneous source 
distribution ($\setminus$ hatches)
and the PSCz distribution (/ hatches). The solid and dashed
lines are the results of the analytical calculations for the same two
cases.  The dash-dotted and dash-dot-dot-dotted lines trace 
the mean simulated fluxes for the homogeneous and the PSCz cases. 
(see [7]).}}
 \end{center}
\end{figure}

\section{Conclusion}

Next generation experiments  such as the High Resolution Fly's Eye which
recently started operating, the Pierre Auger Project which is now under
construction,  the proposed  Telescope Array, and
the EUSO and OWL  satellites  will 
significantly improve the data at the extremely-high end of the cosmic
ray spectrum.\cite{revdata} With these observatories a clear
determination of the spectrum and spatial distribution of UHECR
sources is within reach.

The lack of a GZK cutoff should become clear with HiRes and Auger
and most extragalactic Zevatrons may be ruled out. 
 The observed spectrum will distinguish Zevatrons from
new physics models by testing the hardness of the spectrum and the
effect of propagation.  Figure 11 shows how clearly Auger will test the
spectrum in spite of their clustering properties. The cosmography of sources
should also become clear and able to
 discriminate  between plausible populations for UHECR sources. 
The  correlation of arrival directions  for events with energies above
$10^{20}$ eV  with some known structure such as the Galaxy, the
Galactic halo, the Local Group or the Local Supercluster would be key
in differentiating between different models. For instance, a
correlation with the Galactic center and  disk should become apparent
at extremely high energies for the case of young neutron star winds,
while a correlation with the large scale galaxy distribution should
become clear for the case of quasar remnants. If SHRs or monopolonia are
responsible for UHECR production, the arrival directions should correlate
with the dark matter distribution and show the halo asymmetry. For these
signatures to be tested, full sky coverage is essential. Finally,  an
excellent discriminator would be an unambiguous composition
determination  of the primaries. In general, Galactic disk models  invoke
iron nuclei to be consistent with the isotropic distribution, 
extragalactic Zevatrons tend to favor proton primaries, while photon
primaries are more common for early universe relics. The hybrid detector
of the Auger Project should help determine the composition by measuring
the depth of shower maximum and the muon content of the
same shower.  The prospect of testing extremely high energy physics as
well as solving the UHECR mystery awaits improved observations that
should be coming in the next decade with experiments under construction
such as Auger\cite{Auger} or in the planning stages such as the Telescope
Array,\cite{TA} EUSO,\cite{EUSO} and OWL.\cite{OWL}

\section{Acknowledgment}
 
I thank the organizers of the Relativistic Aspects of Nuclear Physics
2000 meeting, in particular, T. Kodama for his hospitality. This work was supported by NSF through grant AST-0071235  and DOE grant
DE-FG0291 ER40606.

\end{document}